\pdfoutput=1
\documentclass[pra,reprint,showpacs,amsmath,amssymb]{revtex4-1}
\usepackage{mathptm}
\usepackage{hyperref}
\usepackage{graphicx}
\usepackage{xspace,xcolor,textcomp}
\usepackage[figure,table]{hypcap}

\definecolor{darkblue}{rgb}{0.459, 0.439, 0.702}
\hypersetup{colorlinks=true,
linkcolor=blue,
filecolor=blue,
citecolor=blue,
urlcolor=blue,
pdfauthor={Matthias Rosenkranz and Weizhu Bao},
pdftitle={Scattering and bound states in two-dimensional anisotropic potentials},
pdfkeywords={bound states, scattering, polar molecules}}

\newcommand{\ie}{\textit{i.e.}\xspace}
\newcommand{\eg}{\textit{e.g.}\xspace}

\newcommand{\bvec}[1]{\ensuremath{\boldsymbol{\mathrm{#1}}}}
\newcommand{\im}{\ensuremath{\mathrm{i}}}
\newcommand{\eu}{\ensuremath{\mathrm{e}}}
\renewcommand{\Im}{\ensuremath{\mathrm{Im}}}
\renewcommand{\Re}{\ensuremath{\mathrm{Re}}}
\DeclareFontFamily{U}{euc}{}
\DeclareFontShape{U}{euc}{m}{n}{<-6>eurm5<6-8>eurm7<8->eurm10}{}%
\DeclareSymbolFont{AMSc}{U}{euc}{m}{n} 
\DeclareMathSymbol{\umu}{\mathord}{AMSc}{"16}

\begin{document}
\title{Scattering and bound states in two-dimensional anisotropic potentials}
\author{Matthias Rosenkranz}
\author{Weizhu Bao}
\affiliation{Department of Mathematics, National University of
  Singapore, 119076, Singapore}
\date{\today}

\begin{abstract}
  We propose a framework for calculating scattering and bound state
  properties in anisotropic two-dimensional potentials.  Using our
  method, we derive systematic approximations of partial wave phase
  shifts and binding energies.  Moreover, the method is suitable for
  efficient numerical computations.  We calculate the s-wave phase
  shift and binding energy of polar molecules in two layers polarized
  by an external field along an arbitrary direction.  We find that
  scattering depends strongly on their polarization direction and that
  absolute interlayer binding energies are larger than thermal
  energies at typical ultracold temperatures.
\end{abstract}
\maketitle

\section{Introduction}
Anisotropic interactions are at the heart of many physical systems,
such as the atomic nucleus, heteronuclear molecules, or ultracold
dipolar atoms.  The anisotropy has a marked influences on their
scattering and bound state properties.  Additionally, confining a
system to layers often leads to exotic effects and is thought to be a
crucial ingredient for the elusive theory of high-$T_c$
superconductivity~\cite{LeeNagWen06}.

Heteronuclear molecules are a prime candidate for studying low-energy
anisotropic scattering in a controlled manner.  Notably, they possess
a large permanent dipole moment.  Further, several species have now
been cooled to ultracold
temperatures~\cite{NiOspMir08,*DeiGroRep08,*VoiTagCos09}.  These
conditions give rise to an anisotropic dipole-dipole interaction
(DDI), which can be controlled by external
fields~\cite{Bar08,*LahMenSan09,*PupMicBue08}.  For example, an
electric field polarizes the molecules.  However, if the DDI dominates
the three-dimensional (3D) dynamics, the gas becomes unstable
\cite{LahMetFro08}.  Confining the molecules to a two-dimensional (2D)
layer stabilizes the gas for perpendicular
polarization~\cite{Fis06,*KocLahMet08}.  Perpendicularly polarized
molecules in two parallel layers give rise to interlayer bound states
and scattering across
layers~\cite{KlaPikSan10,BarMicRon11,ShiWan09,*PotBerWan10}.  Varying
the polarization direction with an external field controls the
anisotropy of their interaction.  However, only a very recent work has
studied the influence of this DDI anisotropy on bound
states~\cite{VolFedJen11,*VolZinFed11}.

In this paper, we propose a general framework for calculating
scattering phase shifts and binding energies for a wide class of 2D
potentials including anisotropic potentials and the special case of
vanishing potential volume $\int d^2\bvec r V(\bvec r)=0$.  Our method
and calculation extend the results in a central 2D potential by
\citet{KlaPikSan10} to the anisotropic case.  Bound states and
scattering in other classes of 2D potentials have been studied in the
past~\cite{BolGes84,Sim76}.  On the one hand, our formalism allows for
efficient numerical computations in 2D as it only requires solving a
first-order differential equation.  On the other hand, we present
systematic approximations for scattering phase shifts and binding
energy in anisotropic 2D potentials.  Specifically, we recover the
Born approximation for the phase shifts and an approximation
connecting the binding energy and the total phase shift at low
energies.  As an example, we calculate the interlayer binding energy
and s-wave phase shift of polar molecules in a bilayer at arbitrary
polarization.  The interlayer binding, \eg, of two
\textsuperscript{6}Li\textsuperscript{40}K molecules, is deeper than
thermal energies in typical ultracold experiments over a range of
polarization angles.

We consider two particles interacting via a 2D potential $V(\bvec r)$.
In the center-of-mass frame the scattered state depends on both the
relative vector $\bvec r = r(\cos\varphi, \sin\varphi)$ between the
two particles and the incoming wave vector $\bvec k = k(\cos\xi,
\sin\xi)$.  Consequently, we expand the dimensionless wave function in
partial waves for both vectors: $\Psi(\bvec k, \bvec r) =
\tfrac{1}{2\pi} \sum_{m,n=-\infty}^\infty \tfrac{\psi_{mn}(k,
  r)}{\sqrt r} \eu^{\im m\varphi} \eu^{-\im n\xi}$.  Inserting this
expansion into the radial Schr{\"o}dinger equation of the two
particles results in
\begin{equation}\label{eq:Schroedinger}
  \left[\partial^2_r  - \frac{m^2 - 1/4}{r^2} + k^2 \right] \psi_{mn}(k,
  r) = \sum_{m'} V_{mm'}(r) \psi_{m'n}(k, r).
\end{equation}
Here, $k = \sqrt{E 2\mu \Delta^2/\hbar^2}$ is the collision momentum
at energy $E$, $\mu$ is the reduced mass, and $\Delta$ is the unit of
length.  The matrix elements of the potential are $V_{mm'}(r) =
\frac{\mu\Delta^2}{\pi \hbar^2} \int_0^{2\pi} d\varphi \eu^{\im\varphi
  (m-m')} V(\bvec r)$.

Different physical solutions of Eq.~\eqref{eq:Schroedinger}, such as
scattering or bound states, fulfill different boundary conditions.
For our proposed formalism, we expand all physical states $\psi_{mn}$
in a set of \emph{regular} solutions of Eq.~\eqref{eq:Schroedinger},
$\phi_{mn}$, with well-defined boundary conditions.  Specifically, we
require that these regular solutions behave at the origin as $\lim_{r
  \rightarrow 0} \phi_{mn}(k, r)/j_n(kr) = \delta_{mn}$.  Here, we
have defined the scaled Bessel functions $j_m(x) = \sqrt{\pi x/2}
J_m(x)$ and $y_m(x) = \sqrt{\pi x/2} Y_m(x)$, where $J_m$ and $Y_m$
are Bessel functions of the first and second kind, respectively.  For
2D scattering, these scaled Bessel functions are the equivalent of the
Riccati-Bessel functions familiar from 3D scattering~\cite{New82}.
Our choice of the boundary condition fixes the freedom in the
expansion coefficients of the regular solution and its
derivative~\cite{RakSof98}.

We assume $\lim_{r\rightarrow 0} r^2 V_{mn}(r) = 0$ and
$\lim_{r\rightarrow\infty} rV_{mn}(r) = 0$.  Because of the latter, at
large distances solutions of Eq.~\eqref{eq:Schroedinger} are
proportional to the scaled Hankel functions $h_m^{\pm}(x) = j_m(x) \pm
\im y_m(x)$.  This motivates us to expand the regular solutions as
\begin{equation}\label{eq:phi_mn}
  \phi_{mn}(k, r) = \frac{1}{2} [h_m^{+}(kr) f_{mn}^{+}(k, r) +
  h_m^{-}(kr) f_{mn}^{-}(k, r)].
\end{equation}
We insert this expansion into Eq.~\eqref{eq:Schroedinger} and require
that $h_m^{+}(kr) \partial_r f_{mn}^{+}(k, r) + h_m^{-}(kr) \partial_r
f_{mn}^{-}(k, r) = 0$.  This condition reduces the Schr{\"o}dinger
equation to the first order equation
\begin{equation}\label{eq:partial-f_mn}
  \begin{split}
    \partial_r f_{mn}^\pm(k, r) &= \pm \frac{h_m^{\mp}(kr)}{\im k}
    \sum_{m'} V_{mm'}(r) \phi_{m'n}(k, r)
  \end{split}
\end{equation}
for the coefficients $f_{mn}^\pm$.  The formal solution of
Eq.~\eqref{eq:partial-f_mn} is
\begin{equation}\label{eq:F}
  f^\pm_{mn}(k, r) = \delta_{mn} \pm
  \frac{1}{\im k} \int_0^r dr' h^+_m(kr')\sum_{m'} V_{mm'}(r')
  \phi_{m'n}(k, r').
\end{equation}
Here, we have fixed the integration constants so that $f_{mn}^\pm(k,
\infty)$ exhibits the correct high-energy behavior~\cite{RakSof98}.

\section{Bound states}
A weak 2D potential supports bound states if $\int d^2\bvec r V(\bvec
r) \leq 0$~\cite{Sim76}.  Using the coefficients $f_{mn}^\pm$ we
locate bound states in the following way.  For $\Im(k) > 0$ [$\Im(k) <
0$] $f_{mn}^-(k, r)$ [$f_{mn}^+(k, r)$] converges as $r \rightarrow
\infty$ because the right-hand side of Eq.~\eqref{eq:partial-f_mn}
vanishes sufficiently quickly for the assumed long-range behavior of
the potential~\cite{RakSof98}.  The functions $f_{mn}^\pm(k) =
\lim_{r\rightarrow\infty} f_{mn}^\pm(k,
r)|_{\Im(k)\genfrac{}{}{0pt}{}{<}{>}0}$ are the Jost functions
familiar from general scattering theory~\cite{New82,RakSof98}.
Introducing the matrices $F^\pm(k) = [f_{mn}^\pm(k)]$ we locate bound
states by finding momenta $k_b = \im |k_b|$ on the positive imaginary
axis with vanishing determinant, \ie,
\begin{equation}\label{eq:det-F}
  \det [F^-(k_b)] = 0 \Leftrightarrow \sum_n f_{mn}^-(k_b)
  c_n(k_b) = 0.
\end{equation}
Iff the determinant of the Jost matrix $F^-$ vanishes, then there
exists a nonvanishing set of coefficients $c_n$ fulfilling the
right-hand side of Eq.~\eqref{eq:det-F}.  Moreover, at large distances
the solutions $u_m(k_b, r) = \sum_n \phi_{mn}(k_b, r) c_n(k_b)$ vanish
exponentially because $u_m(k_b, r) = \mathcal{O}[h_m^+(k_br)] =
\mathcal{O}[\eu^{-|k_b|r}] \xrightarrow[r\rightarrow\infty]{} 0$.
Therefore, $u_m(k_b, r)$ describes a bound state.  Its binding energy
is $E_b = -|k_b|^2$ in units of $\hbar^2/2\mu\Delta^2$.

Now we are going to derive an approximate expression for the binding
energy in a weak anisotropic 2D potential $V(\bvec r) = V_0 \overline
V(\bvec r)$, where $V_0$ characterizes the strength of the potential.
First, we introduce the explicit expression for the regular solutions
$\phi_{mn}(k, r) = j_n(kr)\delta_{mn} - \int_0^r dr' g_m(k, r, r')
\sum_{m'} V_{mm'}(r) \phi_{m'n}(k, r')$, where $g_m(k, r, r') = k^{-1}
[j_m(kr) y_m(kr') - j_m(kr') y_m(kr)]$ is the free Green's function of
Eq.~\eqref{eq:Schroedinger}~\cite{New86}.  Furthermore, the regular
solutions of the Schr\"odinger Eq.~\eqref{eq:Schroedinger} for $k=0$
are $\phi_{mn}(r) = r^{n+1/2}\delta_{mn} - \int_0^r dr' g_m(r, r')
\sum_{m'} V_{mm'}(r) \phi_{m'n}(r')$, where $g_0(r, r') = \sqrt{rr'}
\ln(r'/r)$ and $g_{m\neq 0}(r, r') = (\sqrt{rr'}/2m) [(r'/r)^m -
(r/r')^m]$ are corresponding free Green's functions.  Now
$\sqrt\pi(k/2)^{n+1/2}\phi_{mn}(r)/n!$ is the expansion of
$\phi_{mn}(k, r)$ to leading order in $k$.  In Eq.~\eqref{eq:F} we
replace $\phi_{mn}(k, r)$ and the scaled Hankel function by their
respective leading order terms.  Then we insert the asymptotic form of
this approximation for $f_{mn}^-(k)$ into the left-hand side of
Eq.~\eqref{eq:det-F}.  The result is $\det[F^-(k)] = A[\ln(k) + \gamma
- \ln(2) -\im\pi/2] + 1 - B = 0$, where $\gamma$ is the
Euler-Mascheroni constant.  Solving this for $k$ we find the binding
energy
\begin{equation}\label{eq:E_b}
  E_b = -4\eu^{-2\gamma - 2\frac{1-B}{A}}.
\end{equation}
We obtain the power series $A = \sum_{j=1}^\infty A^{(j)}$ and $B =
\sum_{j=1}^\infty B^{(j)}$ by expanding $\phi_{mn}(r)$ into a power
series of the potential strength $V_0$, where $j$ indicates the power
of $V_0$.  For a general (possibly anisotropic) 2D potential the
leading terms of these series, including all partial waves, are given
by
\begin{align}
  A^{(1)} &= -\int_0^\infty dr r V_{00}(r),\label{eq:A_1}\\
  A^{(2)} &= \int_0^\infty dr r V_{00}(r) \int_0^r dr' r'
  \ln\left(\frac{r'}{r}\right) V_{00}(r')\label{eq:A_2}\\
  &\quad - \sum_{m\neq 0} \frac{1}{2m} \int_0^\infty dr r\ V_{00}(r)
  \int_0^\infty dr r V_{mm}(r)\notag\\
  &\quad + \sum_{m\neq 0} \int_0^\infty dr \sqrt r V_{0m}(r)
  u_{m0}^{(1)}(r),\notag\\
  B^{(1)} &= \int_0^\infty dr r \ln(r) V_{00}(r) - \sum_{m\neq 0}
  \frac{1}{2m} \int_0^\infty dr r V_{mm}(r),\label{eq:B_1}\\
  B^{(2)} &= -\int_0^\infty dr r \ln(r) V_{00}(r) \int_0^r dr'
  r' \ln\left(\frac{r'}{r}\right) V_{00}(r')\label{eq:B_2}\\
  &\quad + \sum_{m\neq 0} \frac{1}{2m} \int_0^\infty dr r \ln(r)
  V_{00}(r) \int_0^\infty dr r V_{mm}(r)\notag\\
  &\quad - \sum_{m\neq 0} \int_0^\infty dr \sqrt r \ln(r) V_{0m}(r)
  u_{m0}^{(1)}(r)\notag\\
  &\quad + \sum_{m\neq 0} \frac{1}{2m} \int_0^\infty dr r^{1-m}
  V_{m0}(r) \int_0^r dr' r'^{1+m} \ln\left(\frac{r'}{r}\right)
  V_{0m}(r')\notag\\
  &\quad + \sum_{m\neq 0} \frac{1}{2m} \int_0^\infty dr r^{1/2-m}
  [V_{mm}(r) w_m(r) + \sum_{\substack{n\neq 0\\ n\neq m}} V_{mn}(r)
  u_{nm}^{(1)}(r)].\notag
\end{align}
Here, $u_{nm}^{(1)}(r) = \int_0^r dr' r'^{1/2+m} g_n(r, r') V_{nm}(r')
+ (1/2n) r^{1/2+n} \int_0^\infty dr' r'^{1+m-n} V_{nm}(r')$ and
$w_{m}(r) = \int_0^r dr' r'^{1/2+m} g_m(r, r') V_{mm}(r') -
\sum_{n\neq 0, m} (1/2n) r^{1/2+m} \int_0^\infty dr'
r' V_{nn}(r')$.  We can calculate systematically higher order terms by
expanding $\det[F^-(k)]$ to higher orders in $V_0$.  Very recently,
\citet{VolFedJen11} also found such an expansion of the binding energy
by solving the 2D Schr{\" o}dinger equation directly for bound
states.

For concreteness let us now consider two polarized dipoles trapped in
two parallel layers separated by a distance $\Delta$.  We define the
$z$ axis to be perpendicular and the $x$-$y$ plane to be parallel to
the layers.  Without loss of generality, we assume that the dipoles are
polarized within the $x$-$z$ plane at an angle $\vartheta$ from the
$z$ axis.  They interact via the interlayer potential
\begin{equation}\label{eq:V}
  V(\bvec r) = V_0 \frac{\bvec r^2 + 1 - 3 (r\cos
    \varphi\sin\vartheta + \cos\vartheta)^2}{(\bvec r^2 + 1)^{5/2}}.
\end{equation}
Here, $\bvec r = r (\cos\varphi, \sin\varphi)$ is the dimensionless
projected vector between the dipoles in polar coordinates (in units of
$\Delta$) and $V_0 = \mu d^2/2\pi\hbar^2\epsilon_0\Delta$ is the
interaction strength in units of $\hbar^2/2\mu\Delta^2$, with
$\epsilon_0$ the electric constant and $d$ the electric dipole moment
(for magnetic dipoles $V_0 = \mu\mu_0 \mu_d^2/2\pi\hbar^2\Delta$ with
$\mu_0$ the magnetic constant and $\mu_d$ the magnetic dipole moment).
The potential fulfills $A^{(1)} = 0$ and $\int d^2\bvec r V(\bvec r) =
0$ so that at least one bound state exists for all polarization
angles.  For $\vartheta=0$, this potential reduces to the central case
discussed in Refs.~\cite{KlaPikSan10, BarMicRon11}.  Using
Eq.~\eqref{eq:E_b} and expanding $A$ and $B$ to fourth and second
order, respectively, we recover the binding energy for perpendicular
polarization given in Ref.~\cite{BarMicRon11}.

\begin{figure}
  \centering
  \includegraphics[width=.95\linewidth]{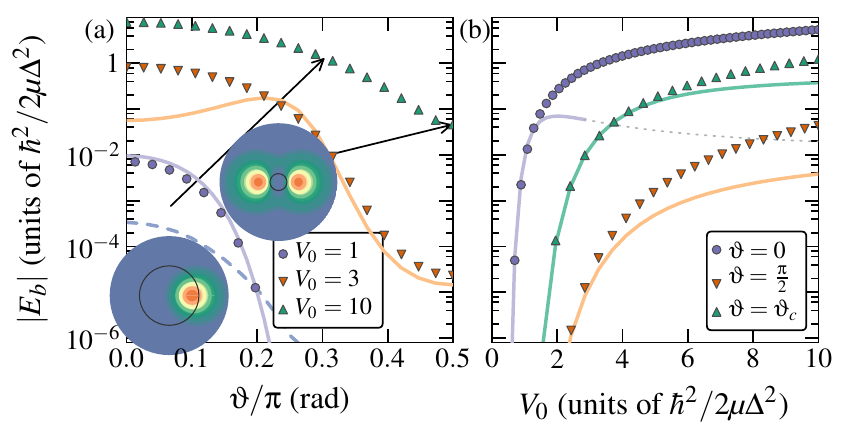}
  \caption{(Color online) Binding energy of lowest interlayer bound
    states in the bilayer dipolar potential Eq.~\eqref{eq:V} as a
    function of (a) the polarization angle from the symmetry axis and
    (b) interaction strength $V_0$.  The symbols mark numerical
    results, the dashed line an approximation in Ref.~\cite{Sim76} for
    $V_0=1$, and solid lines the corresponding approximation
    Eqs.~\eqref{eq:E_b}-\eqref{eq:B_2}.  The latter approximation
    remains valid up to moderate $V_0$ as long as $|E_b| \ll 1$.  The
    insets show densities of the bound states at the indicated
    polarization angles and $V_0=10$.  The dark circles indicate the
    radius of one layer distance $\Delta$.}\label{fig:bound-states}
\end{figure}

In Fig.~\ref{fig:bound-states} we plot the binding energy of the
lowest lying interlayer bound states in the potential~\eqref{eq:V} as
a function of the polarization angle and interaction strength.  Their
binding is strongest for perpendicular polarization and weakest for
parallel polarization.  We observe that the weak potential
approximation for the binding energy,
Eqs.~\eqref{eq:E_b}-\eqref{eq:B_2}, remains valid up to moderate
potential strengths $V_0 \alt 3$ as long as the binding energy is
sufficiently small, $|E_b| \ll 1$.  In contrast, the approximation in
Ref.~\cite{Sim76}, $|E_b| \sim \exp(1/c)$ with $c = (1/8\pi^2) \int
d^2\bvec r d^2\bvec r' V(\bvec r) \ln|\bvec r - \bvec r'| V(\bvec
r')$, describes the angle dependence of the binding energy only
poorly.  Our numerical computations are based on Netlib's
\textsc{zvode} solver, and we provide an explicit Jacobian for improved
stability.  We include partial waves up to sixth order.

As an example, we consider bosonic
\textsuperscript{6}Li\textsuperscript{40}K molecules separated by
$\Delta = 200\,\mathrm{nm}$.  Then the energy scale in
Fig.~\ref{fig:bound-states} is $\hbar^2/2\mu\Delta^2 \simeq
1.2\,\umu\mathrm{K}$ (units of $k_B$) and $V_0 \simeq 9.5$
(cf. $V_0=10$ in Fig.~\ref{fig:bound-states}), with $k_B$ the
Boltzmann constant.  Therefore, the interlayer bound state of LiK
molecules should persist over a wide range of polarization angles
under typical experimental temperatures in the nano-Kelvin regime.
For fermionic \textsuperscript{40}K\textsuperscript{87}Rb molecules we
have $V_0 \simeq 1$ with energy scale $\simeq 220\,\mathrm{nK}$ (units
of $k_B$).  Therefore, it can be seen from
Fig.~\ref{fig:bound-states}(a) that they require much lower
temperatures.  In contrast, for dipolar atoms with a magnetic dipole
moment, $V_0 \ll 1$ (\eg, \textsuperscript{52}Cr).  Interlayer dipoles
with $V_0 \ll 1$ bind too weakly for all polarization angles to be
stable at reasonable external parameters.

The peak of the bound state wave function shifts along the $x$ axis as
the polarization direction changes from perpendicular to parallel
[cf. insets in Fig.~\ref{fig:bound-states}(a)].  This is because the
attractive dipole term in the DDI dominates over the quadrupole term.
However, for $\vartheta > 0.47\pi$, the quadrupole term dominates and
two peaks develop.  For parallel polarization, the dipole term
vanishes and the two peaks become symmetric.  In
Fig.~\ref{fig:bound-states}(a) the distance between these peaks is
$\simeq 5\Delta$, typically on the order of $\umu\mathrm{m}$.  If the
stability requirements are fulfilled, this could simplify the creation
of interlayer bound states at parallel polarization because of their
greater overlap than states at perpendicular polarization.  The
parallel polarization bound state can be distinguished from the mainly
perpendicular one in a time of flight measurement.  The time-of-flight
image reflects the double peak in an asymmetric momentum distribution.

\section{Scattering}
Next we focus on the 2D scattering problem, for which $k$ is real.
Then the functions $f_{mn}^\pm(k, r)$ attain the finite limit
$f_{mn}^\pm(k) = \lim_{r\rightarrow\infty} f_{mn}^\pm(k, r)$ because
the right-hand side of Eq.~\eqref{eq:partial-f_mn} vanishes at $r
\rightarrow \infty$.  In order to capture the mixing of different
partial waves, it is necessary to calculate the full S-matrix.  We
express the partial wave components of the scattering solution as a
linear combination $\psi_{mn}(k, r) = \sum_{m'} \phi_{mm'}(k, r)
c_{m'n}(k)$ of the regular solutions, Eq.~\eqref{eq:phi_mn}.
Asymptotically, we replace $f_{mm'}^\pm(k, r)$ in $\phi_{mm'}(k, r)$
with the Jost functions $f_{mm'}^\pm(k)$.  On the other hand, the
general asymptotic scattering wave function in 2D is $\Psi(\bvec k,
\bvec r) \rightarrow [\eu^{\im \bvec k \cdot \bvec r} + a(\bvec k,
\bvec r) \frac{\eu^{\im kr}}{\sqrt r}] / 2\pi$, where $a(\bvec k,
\bvec r)$ is the scattering amplitude.  We match the two asymptotic
expressions for the scattering wave functions by expanding the first
exponential in $\Psi(\bvec k, \bvec r)$ in terms of scaled Hankel
functions.  This way, we extract the S-matrix
\begin{equation}\label{eq:S}
  S(k) = F^+(k) [F^-(k)]^{-1}
\end{equation}
and the coefficients $c_{mn}(k) = [F^-(k)]^{-1}_{mn}$.  If the
potential $V(\bvec r)$ is central, $S(k)$ is diagonal, with elements
$\eu^{2\im\delta_m(k)}$ and $\delta_m(k)$ the $m$-th partial wave
phase shift.

Let us now derive approximate expressions for the phase shifts of
anisotropic 2D scattering at low energies.  First we introduce an
iterative solution for the coefficients $f_{mn}^\pm(k, r) =
\sum_{j=0}^\infty f_{mn}^{\pm (j)}(k, r)$ as a power series in the
potential strength $V_0$.  From Eq.~\eqref{eq:F} and the expression
for $\phi_{mn}(k, r)$ we obtain $f_{mn}^{\pm (j+1)}(k, r) =
\pm\tfrac{1}{\im k} \int_0^r dr' h_m^\mp(kr') \sum_{m'} V_{mm'}(r')
\phi_{m'n}^{(j)}(k, r')$ and $f_{mn}^{\pm (0)}(k, r) = \delta_{mn}$,
with $\phi_{mn}^{(j+1)}(k, r) = -\int_0^r dr' g_m(k, r, r') \sum_{m'}
V_{mm'}(r') \phi_{m'n}^{(j)}(k, r')$ and $\phi_{mn}^{(0)}(k, r) =
j_n(kr)\delta_{mn}$.  In the remainder of this section, we consider
low scattering energies such that only up to two partial waves $\ell$
and $\ell'$ dominate the properties of the S-matrix.  The phase shifts
are given by $\tan 2\delta_\ell(k) = \Im S_{\ell\ell}/\Re
S_{\ell\ell}$, where $S_{\ell\ell}$ is a diagonal matrix element of
$S$ and $\delta_{\ell'}$ is obtained by replacing $\ell \rightarrow
\ell'$.  By inserting Eq.~\eqref{eq:S} we obtain the phase shifts from
the Jost matrix $F^-$ since $F^+ = \overline{F^-}$ for real $k$.  Then
\begin{equation}\label{eq:tan-delta-frac}
  \tan 2\delta_\ell(k) \simeq 2\frac{\Im s_{\ell\ell}}{\Re s_{\ell\ell}}
\end{equation}
with $s_{\ell\ell} = S_{\ell\ell}|\det(F^-)|^2$.  Using terms
$f_{mn}^{-(j)}$ up to second order and expanding $s_{\ell\ell}$ to
second order in $V_0$ we find $\Im(s_{\ell\ell}) = -I^{(1)}_{j_\ell
  j_\ell} + I^{(1)}_{j_\ell j_\ell} I^{(1)}_{y_\ell j_\ell} -
I^{(1)}_{j_\ell j_{\ell'}} I^{(1)}_{y_{\ell'} j_\ell} + 2
I^{(1)}_{j_\ell j_\ell} I^{(1)}_{y_{\ell'}j_{\ell'}} +
I^{(2)}_{j_\ell},$ $\Re(s_{\ell\ell}) = 1 + \sum_{m=\ell, \ell'}
[-2I^{(1)}_{y_m j_m} + 2 I^{(2)}_{y_m} + I^{(1)^2}_{j_{m}j_{m}} +
I^{(1)^2}_{y_m j_m}] - 2I^{(1)^2}_{j_\ell j_\ell} - 2 I^{(1)}_{y_\ell
  j_{\ell'}} I^{(1)}_{y_{\ell'} j_\ell} + 4 I^{(1)}_{y_\ell j_\ell}
I^{(1)}_{y_{\ell'} y_{\ell'}}$, $I_{p_m q_n}^{(1)}(k) = (1/k)
\int_0^\infty dr p_m(kr) V_{mn}(r) q_n(kr)$, $I^{(2)}_{p_n}(k) = (1/k)
\int_0^\infty dr p_n(kr) \sum_{m} V_{n m}(r) \int_0^r dr' g_m(k, r,
r') V_{mn}(r') j_n(kr')$ and $p_m$, $q_n$ stand in for scaled Bessel
functions of order $m$ and $n$, respectively.  The anisotropy of the
potential enters the phase shift through mixing terms, such as
$I^{(1)}_{j_\ell j_{\ell'}}$, and the sum in $I^{(2)}_{p_n}$.  These
terms are absent for a central potential so we recover a result in
Ref.~\cite{KlaPikSan10}.  Furthermore, for small $V_0$ we expand
Eq.~\eqref{eq:tan-delta-frac} to second order in $V_0$:
\begin{equation}\label{eq:tan-delta-Born}
  \tan 2\delta_\ell(k) \simeq -2 I_{j_\ell j_\ell}^{(1)}(k) +
  2 [I^{(2)}_{\ell} - I^{(1)}_{j_\ell j_\ell} I^{(1)}_{y_\ell
    j_\ell} - I^{(1)}_{j_\ell j_{\ell'}} I^{(1)}_{y_{\ell'}
    j_\ell}].
\end{equation}
Thus, we recover a second-order Born approximation of the phase shift
from our general formalism.  We find a further approximation by
considering the sum of all partial waves at small energies.  For any
S-matrix, $\cot\sum_\ell\delta_\ell(k) = \Re(\det S)/\Im(\det S) =
-\Re(\det F^-)/\Im(\det F^-)$.  As in the preceding section, we expand
$\det F^-$ around $k=0$ and use the fact that $k$ is real to obtain
\begin{equation}\label{eq:tan-deltas}
  \cot\sum_\ell \delta_\ell(k) \simeq \frac{1}{\pi} \ln\left(
      \frac{k^2}{|E_b|} \right).
\end{equation}
Here, $E_b$ is the binding energy, Eq.~\eqref{eq:E_b}.

\begin{figure}
  \centering
  \includegraphics[width=.95\linewidth]{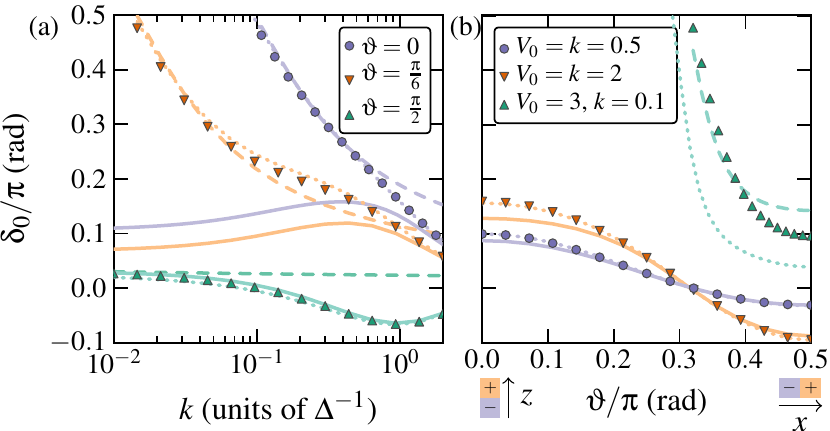}
  \caption{(Color online) S-wave phase shift for anisotropic
    interlayer scattering of dipoles as a function of (a) the
    collision momentum for $V_0=1$ (\eg, KRb molecules at
    $\Delta=200\,\mathrm{nm}$) and (b) the polarization angle.  The
    symbols mark numerical results, solid lines the second-order Born
    approximation Eq.~\eqref{eq:tan-delta-Born}, dashed lines
    Eq.~\eqref{eq:tan-deltas}, and dotted lines
    Eq.~\eqref{eq:tan-delta-frac} to second order in $f_{mn}^{-(j)}$.
    The weak potential approximation~\eqref{eq:tan-delta-frac}
    describes scattering well from moderately small collision momenta
    and up to moderate potential strengths at all polarizations. The
    bound-state approximation~\eqref{eq:tan-deltas} is only accurate
    at very low momenta and predominantly perpendicular
    polarization.}\label{fig:scattering}
\end{figure}

In Fig.~\ref{fig:scattering}(a) we plot the s-wave phase shift of two
dipoles with $V_0=1$ interacting across two layers as a function of
the collision momentum $k$, \eg,
\textsuperscript{40}K\textsuperscript{87}Rb with different spins at
$\Delta=200\,\mathrm{nm}$.  The sharp increase of the s-wave phase
shift at $k\rightarrow 0$ ($\vartheta=0, \pi/6$) is a consequence of
the very weakly bound states.  If $\delta_0$ dominates in
Eq.~\eqref{eq:tan-deltas}, a very small binding energy leads to a
phase jump in $\delta_0$ close to $k=0$.  Since the binding energy
decreases strongly with increasing polarization angle,
expression~\eqref{eq:tan-deltas} describes the scattering of mainly
in-plane polarization only at unrealistically small collision
energies.  On the other hand, the weak potential
approximation~\eqref{eq:tan-delta-frac} describes the numerics
excellently at all considered momenta.  It fails at large potential
strengths $V_0\gg 1$ and small momenta.  For large momenta $k\gg 1$
this approximation becomes identical with the Born
approximation~\eqref{eq:tan-delta-Born}.  In
Fig.~\ref{fig:scattering}(b) we observe that the s-wave phase shift
can vary strongly with the polarization angle at small momenta.  The
phase jump is caused by a weakly bound state at $\vartheta\simeq 0.3
\pi$ [see Fig.~\ref{fig:bound-states}(a)].  This variability should be
observable in the scattering of polar molecules.  The bound-state
approximation~\eqref{eq:tan-deltas} describes this behavior
qualitatively even for such moderately large potential strengths as
long as $k \ll 1$.  The difference is mainly due to neglecting higher
order partial waves in Eq.~\eqref{eq:tan-deltas}.  For a small
potential strength $V_0 \ll 1$ our approximation
Eq.~\eqref{eq:tan-delta-frac} describes scattering excellently at all
polarizations.  We find that this approximation remains valid even at
moderately large potential strengths $V_0 \alt 3$ at higher energies
$k^2 \agt 1$.

\section{Conclusions}
We have proposed a framework for calculating scattering and bound
state properties for anisotropic 2D potentials.  Our method
generalizes the Jost formalism known from 3D scattering.  We have
derived systematic approximations for the scattering phase shifts and
binding energy at low to moderate potential strengths.  For weak
potentials we have recovered a second-order Born approximation.  The
central equation~\eqref{eq:partial-f_mn} is also well-suited for
numerical computations.

We have applied our method to polar molecules trapped in a bilayer and
polarized along an arbitrary direction.  We find that absolute
energies of \textsuperscript{6}Li\textsuperscript{40}K interlayer
bound states are larger than their thermal energy in ultracold
experiments even for nonperpendicular polarization.  The s-wave phase
shift of molecules with moderate or large DDI exhibits a strong
dependence on the polarization angle.  These results are important,
\eg, for the BEC-BCS crossover in fermionic polar molecules in
bilayers~\cite{PikKlaShl10,ZinWunPek10}.  Varying the direction of the external
polarizing field should influence the crossover from interlayer pair
condensation to BCS pairing.

We thank Dieter Jaksch for helpful discussions.  This work was
supported by the Academic Research Fund of the Ministry of Education
of Singapore, Grant No. R-146-000-120-112.

\bibliographystyle{apsrev4-1}
\bibliography{dipscatt}

\begin{thebibliography}{23}%
\makeatletter
\providecommand \@ifxundefined [1]{%
 \@ifx{#1\undefined}
}%
\providecommand \@ifnum [1]{%
 \ifnum #1\expandafter \@firstoftwo
 \else \expandafter \@secondoftwo
 \fi
}%
\providecommand \@ifx [1]{%
 \ifx #1\expandafter \@firstoftwo
 \else \expandafter \@secondoftwo
 \fi
}%
\providecommand \natexlab [1]{#1}%
\providecommand \enquote  [1]{``#1''}%
\providecommand \bibnamefont  [1]{#1}%
\providecommand \bibfnamefont [1]{#1}%
\providecommand \citenamefont [1]{#1}%
\providecommand \href@noop [0]{\@secondoftwo}%
\providecommand \href [0]{\begingroup \@sanitize@url \@href}%
\providecommand \@href[1]{\@@startlink{#1}\@@href}%
\providecommand \@@href[1]{\endgroup#1\@@endlink}%
\providecommand \@sanitize@url [0]{\catcode `\\12\catcode `\$12\catcode
  `\&12\catcode `\#12\catcode `\^12\catcode `\_12\catcode `\%12\relax}%
\providecommand \@@startlink[1]{}%
\providecommand \@@endlink[0]{}%
\providecommand \url  [0]{\begingroup\@sanitize@url \@url }%
\providecommand \@url [1]{\endgroup\@href {#1}{\urlprefix }}%
\providecommand \urlprefix  [0]{URL }%
\providecommand \Eprint [0]{\href }%
\providecommand \doibase [0]{http://dx.doi.org/}%
\providecommand \selectlanguage [0]{\@gobble}%
\providecommand \bibinfo  [0]{\@secondoftwo}%
\providecommand \bibfield  [0]{\@secondoftwo}%
\providecommand \translation [1]{[#1]}%
\providecommand \BibitemOpen [0]{}%
\providecommand \bibitemStop [0]{}%
\providecommand \bibitemNoStop [0]{.\EOS\space}%
\providecommand \EOS [0]{\spacefactor3000\relax}%
\providecommand \BibitemShut  [1]{\csname bibitem#1\endcsname}%
\let\auto@bib@innerbib\@empty
\bibitem [{\citenamefont {Lee}\ \emph {et~al.}(2006)\citenamefont {Lee},
  \citenamefont {Nagaosa},\ and\ \citenamefont {Wen}}]{LeeNagWen06}%
  \BibitemOpen
  \bibfield  {author} {\bibinfo {author} {\bibfnamefont {P.~A.}\ \bibnamefont
  {Lee}}, \bibinfo {author} {\bibfnamefont {N.}~\bibnamefont {Nagaosa}}, \ and\
  \bibinfo {author} {\bibfnamefont {X.-G.}\ \bibnamefont {Wen}},\ }\href
  {\doibase 10.1103/RevModPhys.78.17} {\bibfield  {journal} {\bibinfo
  {journal} {Rev. Mod. Phys.}\ }\textbf {\bibinfo {volume} {78}},\ \bibinfo
  {pages} {17} (\bibinfo {year} {2006})}\BibitemShut {NoStop}%
\bibitem [{\citenamefont {Ni}\ \emph {et~al.}(2008)\citenamefont {Ni},
  \citenamefont {Ospelkaus}, \citenamefont {de~Miranda}, \citenamefont {Pe'er},
  \citenamefont {Neyenhuis}, \citenamefont {Zirbel}, \citenamefont
  {Kotochigova}, \citenamefont {Julienne}, \citenamefont {Jin},\ and\
  \citenamefont {Ye}}]{NiOspMir08}%
  \BibitemOpen
  \bibfield  {author} {\bibinfo {author} {\bibfnamefont {K.-K.}\ \bibnamefont
  {Ni}}, \bibinfo {author} {\bibfnamefont {S.}~\bibnamefont {Ospelkaus}},
  \bibinfo {author} {\bibfnamefont {M.~H.~G.}\ \bibnamefont {de~Miranda}},
  \bibinfo {author} {\bibfnamefont {A.}~\bibnamefont {Pe'er}}, \bibinfo
  {author} {\bibfnamefont {B.}~\bibnamefont {Neyenhuis}}, \bibinfo {author}
  {\bibfnamefont {J.~J.}\ \bibnamefont {Zirbel}}, \bibinfo {author}
  {\bibfnamefont {S.}~\bibnamefont {Kotochigova}}, \bibinfo {author}
  {\bibfnamefont {P.~S.}\ \bibnamefont {Julienne}}, \bibinfo {author}
  {\bibfnamefont {D.~S.}\ \bibnamefont {Jin}}, \ and\ \bibinfo {author}
  {\bibfnamefont {J.}~\bibnamefont {Ye}},\ }\href {\doibase
  10.1126/science.1163861} {\bibfield  {journal} {\bibinfo  {journal}
  {Science}\ }\textbf {\bibinfo {volume} {322}},\ \bibinfo {pages} {231}
  (\bibinfo {year} {2008})}\BibitemShut {NoStop}%
\bibitem [{\citenamefont {Deiglmayr}\ \emph {et~al.}(2008)\citenamefont
  {Deiglmayr}, \citenamefont {Grochola}, \citenamefont {Repp}, \citenamefont
  {M{\"o}rtlbauer}, \citenamefont {Gl{\"u}ck}, \citenamefont {Lange},
  \citenamefont {Dulieu}, \citenamefont {Wester},\ and\ \citenamefont
  {Weidem{\"u}ller}}]{DeiGroRep08}%
  \BibitemOpen
  \bibfield  {author} {\bibinfo {author} {\bibfnamefont {J.}~\bibnamefont
  {Deiglmayr}}, \bibinfo {author} {\bibfnamefont {A.}~\bibnamefont {Grochola}},
  \bibinfo {author} {\bibfnamefont {M.}~\bibnamefont {Repp}}, \bibinfo {author}
  {\bibfnamefont {K.}~\bibnamefont {M{\"o}rtlbauer}}, \bibinfo {author}
  {\bibfnamefont {C.}~\bibnamefont {Gl{\"u}ck}}, \bibinfo {author}
  {\bibfnamefont {J.}~\bibnamefont {Lange}}, \bibinfo {author} {\bibfnamefont
  {O.}~\bibnamefont {Dulieu}}, \bibinfo {author} {\bibfnamefont
  {R.}~\bibnamefont {Wester}}, \ and\ \bibinfo {author} {\bibfnamefont
  {M.}~\bibnamefont {Weidem{\"u}ller}},\ }\href {\doibase
  10.1103/PhysRevLett.101.133004} {\bibfield  {journal} {\bibinfo  {journal}
  {Phys. Rev. Lett.}\ }\textbf {\bibinfo {volume} {101}},\ \bibinfo {pages}
  {133004} (\bibinfo {year} {2008})}\BibitemShut {NoStop}%
\bibitem [{\citenamefont {Voigt}\ \emph {et~al.}(2009)\citenamefont {Voigt},
  \citenamefont {Taglieber}, \citenamefont {Costa}, \citenamefont {Aoki},
  \citenamefont {Wieser}, \citenamefont {H{\"a}nsch},\ and\ \citenamefont
  {Dieckmann}}]{VoiTagCos09}%
  \BibitemOpen
  \bibfield  {author} {\bibinfo {author} {\bibfnamefont {A.-C.}\ \bibnamefont
  {Voigt}}, \bibinfo {author} {\bibfnamefont {M.}~\bibnamefont {Taglieber}},
  \bibinfo {author} {\bibfnamefont {L.}~\bibnamefont {Costa}}, \bibinfo
  {author} {\bibfnamefont {T.}~\bibnamefont {Aoki}}, \bibinfo {author}
  {\bibfnamefont {W.}~\bibnamefont {Wieser}}, \bibinfo {author} {\bibfnamefont
  {T.~W.}\ \bibnamefont {H{\"a}nsch}}, \ and\ \bibinfo {author} {\bibfnamefont
  {K.}~\bibnamefont {Dieckmann}},\ }\href {\doibase
  10.1103/PhysRevLett.102.020405} {\bibfield  {journal} {\bibinfo  {journal}
  {Phys. Rev. Lett.}\ }\textbf {\bibinfo {volume} {102}},\ \bibinfo {pages}
  {020405} (\bibinfo {year} {2009})}\BibitemShut {NoStop}%
\bibitem [{\citenamefont {Baranov}(2008)}]{Bar08}%
  \BibitemOpen
  \bibfield  {author} {\bibinfo {author} {\bibfnamefont {M.~A.}\ \bibnamefont
  {Baranov}},\ }\href {\doibase 10.1016/j.physrep.2008.04.007} {\bibfield
  {journal} {\bibinfo  {journal} {Phys. Rep.}\ }\textbf {\bibinfo {volume}
  {464}},\ \bibinfo {pages} {71} (\bibinfo {year} {2008})}\BibitemShut
  {NoStop}%
\bibitem [{\citenamefont {Lahaye}\ \emph {et~al.}(2009)\citenamefont {Lahaye},
  \citenamefont {Menotti}, \citenamefont {Santos}, \citenamefont {Lewenstein},\
  and\ \citenamefont {Pfau}}]{LahMenSan09}%
  \BibitemOpen
  \bibfield  {author} {\bibinfo {author} {\bibfnamefont {T.}~\bibnamefont
  {Lahaye}}, \bibinfo {author} {\bibfnamefont {C.}~\bibnamefont {Menotti}},
  \bibinfo {author} {\bibfnamefont {L.}~\bibnamefont {Santos}}, \bibinfo
  {author} {\bibfnamefont {M.}~\bibnamefont {Lewenstein}}, \ and\ \bibinfo
  {author} {\bibfnamefont {T.}~\bibnamefont {Pfau}},\ }\href {\doibase
  10.1088/0034-4885/72/12/126401} {\bibfield  {journal} {\bibinfo  {journal}
  {Rep. Prog. Phys.}\ }\textbf {\bibinfo {volume} {72}},\ \bibinfo {pages}
  {126401} (\bibinfo {year} {2009})}\BibitemShut {NoStop}%
\bibitem [{\citenamefont {Pupillo}\ \emph {et~al.}(2008)\citenamefont
  {Pupillo}, \citenamefont {Micheli}, \citenamefont {B{\"u}chler},\ and\
  \citenamefont {Zoller}}]{PupMicBue08}%
  \BibitemOpen
  \bibfield  {author} {\bibinfo {author} {\bibfnamefont {G.}~\bibnamefont
  {Pupillo}}, \bibinfo {author} {\bibfnamefont {A.}~\bibnamefont {Micheli}},
  \bibinfo {author} {\bibfnamefont {H.~P.}\ \bibnamefont {B{\"u}chler}}, \ and\
  \bibinfo {author} {\bibfnamefont {P.}~\bibnamefont {Zoller}},\ }\href@noop {}
  {\  (\bibinfo {year} {2008})},\ \Eprint {http://arxiv.org/abs/0805.1896}
  {arXiv:0805.1896} \BibitemShut {NoStop}%
\bibitem [{\citenamefont {Lahaye}\ \emph {et~al.}(2008)\citenamefont {Lahaye},
  \citenamefont {Metz}, \citenamefont {Fr{\"o}hlich}, \citenamefont {Koch},
  \citenamefont {Meister}, \citenamefont {Griesmaier}, \citenamefont {Pfau},
  \citenamefont {Saito}, \citenamefont {Kawaguchi},\ and\ \citenamefont
  {Ueda}}]{LahMetFro08}%
  \BibitemOpen
  \bibfield  {author} {\bibinfo {author} {\bibfnamefont {T.}~\bibnamefont
  {Lahaye}}, \bibinfo {author} {\bibfnamefont {J.}~\bibnamefont {Metz}},
  \bibinfo {author} {\bibfnamefont {B.}~\bibnamefont {Fr{\"o}hlich}}, \bibinfo
  {author} {\bibfnamefont {T.}~\bibnamefont {Koch}}, \bibinfo {author}
  {\bibfnamefont {M.}~\bibnamefont {Meister}}, \bibinfo {author} {\bibfnamefont
  {A.}~\bibnamefont {Griesmaier}}, \bibinfo {author} {\bibfnamefont
  {T.}~\bibnamefont {Pfau}}, \bibinfo {author} {\bibfnamefont {H.}~\bibnamefont
  {Saito}}, \bibinfo {author} {\bibfnamefont {Y.}~\bibnamefont {Kawaguchi}}, \
  and\ \bibinfo {author} {\bibfnamefont {M.}~\bibnamefont {Ueda}},\ }\href
  {http://link.aps.org/doi/10.1103/PhysRevLett.101.080401} {\bibfield
  {journal} {\bibinfo  {journal} {Phys. Rev. Lett.}\ }\textbf {\bibinfo
  {volume} {101}},\ \bibinfo {pages} {080401} (\bibinfo {year}
  {2008})}\BibitemShut {NoStop}%
\bibitem [{\citenamefont {Fischer}(2006)}]{Fis06}%
  \BibitemOpen
  \bibfield  {author} {\bibinfo {author} {\bibfnamefont {U.~R.}\ \bibnamefont
  {Fischer}},\ }\href {\doibase 10.1103/PhysRevA.73.031602} {\bibfield
  {journal} {\bibinfo  {journal} {Phys. Rev. A}\ }\textbf {\bibinfo {volume}
  {73}},\ \bibinfo {pages} {031602} (\bibinfo {year} {2006})}\BibitemShut
  {NoStop}%
\bibitem [{\citenamefont {Koch}\ \emph {et~al.}(2008)\citenamefont {Koch},
  \citenamefont {Lahaye}, \citenamefont {Metz}, \citenamefont {Fr{\"o}hlich},
  \citenamefont {Griesmaier},\ and\ \citenamefont {Pfau}}]{KocLahMet08}%
  \BibitemOpen
  \bibfield  {author} {\bibinfo {author} {\bibfnamefont {T.}~\bibnamefont
  {Koch}}, \bibinfo {author} {\bibfnamefont {T.}~\bibnamefont {Lahaye}},
  \bibinfo {author} {\bibfnamefont {J.}~\bibnamefont {Metz}}, \bibinfo {author}
  {\bibfnamefont {B.}~\bibnamefont {Fr{\"o}hlich}}, \bibinfo {author}
  {\bibfnamefont {A.}~\bibnamefont {Griesmaier}}, \ and\ \bibinfo {author}
  {\bibfnamefont {T.}~\bibnamefont {Pfau}},\ }\href {\doibase 10.1038/nphys887}
  {\bibfield  {journal} {\bibinfo  {journal} {Nat. Phys.}\ }\textbf {\bibinfo
  {volume} {4}},\ \bibinfo {pages} {218} (\bibinfo {year} {2008})}\BibitemShut
  {NoStop}%
\bibitem [{\citenamefont {Klawunn}\ \emph {et~al.}(2010)\citenamefont
  {Klawunn}, \citenamefont {Pikovski},\ and\ \citenamefont
  {Santos}}]{KlaPikSan10}%
  \BibitemOpen
  \bibfield  {author} {\bibinfo {author} {\bibfnamefont {M.}~\bibnamefont
  {Klawunn}}, \bibinfo {author} {\bibfnamefont {A.}~\bibnamefont {Pikovski}}, \
  and\ \bibinfo {author} {\bibfnamefont {L.}~\bibnamefont {Santos}},\ }\href
  {\doibase 10.1103/PhysRevA.82.044701} {\bibfield  {journal} {\bibinfo
  {journal} {Phys. Rev. A}\ }\textbf {\bibinfo {volume} {82}},\ \bibinfo
  {pages} {044701} (\bibinfo {year} {2010})}\BibitemShut {NoStop}%
\bibitem [{\citenamefont {Baranov}\ \emph {et~al.}(2011)\citenamefont
  {Baranov}, \citenamefont {Micheli}, \citenamefont {Ronen},\ and\
  \citenamefont {Zoller}}]{BarMicRon11}%
  \BibitemOpen
  \bibfield  {author} {\bibinfo {author} {\bibfnamefont {M.~A.}\ \bibnamefont
  {Baranov}}, \bibinfo {author} {\bibfnamefont {A.}~\bibnamefont {Micheli}},
  \bibinfo {author} {\bibfnamefont {S.}~\bibnamefont {Ronen}}, \ and\ \bibinfo
  {author} {\bibfnamefont {P.}~\bibnamefont {Zoller}},\ }\href {\doibase
  10.1103/PhysRevA.83.043602} {\bibfield  {journal} {\bibinfo  {journal} {Phys.
  Rev. A}\ }\textbf {\bibinfo {volume} {83}},\ \bibinfo {pages} {043602}
  (\bibinfo {year} {2011})}\BibitemShut {NoStop}%
\bibitem [{\citenamefont {Shih}\ and\ \citenamefont {Wang}(2009)}]{ShiWan09}%
  \BibitemOpen
  \bibfield  {author} {\bibinfo {author} {\bibfnamefont {S.-M.}\ \bibnamefont
  {Shih}}\ and\ \bibinfo {author} {\bibfnamefont {D.-W.}\ \bibnamefont
  {Wang}},\ }\href {\doibase 10.1103/PhysRevA.79.065603} {\bibfield  {journal}
  {\bibinfo  {journal} {Phys. Rev. A}\ }\textbf {\bibinfo {volume} {79}},\
  \bibinfo {pages} {065603} (\bibinfo {year} {2009})}\BibitemShut {NoStop}%
\bibitem [{\citenamefont {Potter}\ \emph {et~al.}(2010)\citenamefont {Potter},
  \citenamefont {Berg}, \citenamefont {Wang}, \citenamefont {Halperin},\ and\
  \citenamefont {Demler}}]{PotBerWan10}%
  \BibitemOpen
  \bibfield  {author} {\bibinfo {author} {\bibfnamefont {A.~C.}\ \bibnamefont
  {Potter}}, \bibinfo {author} {\bibfnamefont {E.}~\bibnamefont {Berg}},
  \bibinfo {author} {\bibfnamefont {D.-W.}\ \bibnamefont {Wang}}, \bibinfo
  {author} {\bibfnamefont {B.~I.}\ \bibnamefont {Halperin}}, \ and\ \bibinfo
  {author} {\bibfnamefont {E.}~\bibnamefont {Demler}},\ }\href {\doibase
  10.1103/PhysRevLett.105.220406} {\bibfield  {journal} {\bibinfo  {journal}
  {Phys. Rev. Lett.}\ }\textbf {\bibinfo {volume} {105}},\ \bibinfo {pages}
  {220406} (\bibinfo {year} {2010})}\BibitemShut {NoStop}%
\bibitem [{\citenamefont {Volosniev}\ \emph
  {et~al.}(2011{\natexlab{a}})\citenamefont {Volosniev}, \citenamefont
  {Fedorov}, \citenamefont {Jensen},\ and\ \citenamefont
  {Zinner}}]{VolFedJen11}%
  \BibitemOpen
  \bibfield  {author} {\bibinfo {author} {\bibfnamefont {A.~G.}\ \bibnamefont
  {Volosniev}}, \bibinfo {author} {\bibfnamefont {D.~V.}\ \bibnamefont
  {Fedorov}}, \bibinfo {author} {\bibfnamefont {A.~S.}\ \bibnamefont {Jensen}},
  \ and\ \bibinfo {author} {\bibfnamefont {N.~T.}\ \bibnamefont {Zinner}},\
  }\href {\doibase 10.1103/PhysRevLett.106.250401} {\bibfield  {journal}
  {\bibinfo  {journal} {Phys. Rev. Lett.}\ }\textbf {\bibinfo {volume} {106}},\
  \bibinfo {pages} {250401} (\bibinfo {year} {2011}{\natexlab{a}})}\BibitemShut
  {NoStop}%
\bibitem [{\citenamefont {Volosniev}\ \emph
  {et~al.}(2011{\natexlab{b}})\citenamefont {Volosniev}, \citenamefont
  {Zinner}, \citenamefont {Fedorov}, \citenamefont {Jensen},\ and\
  \citenamefont {Wunsch}}]{VolZinFed11}%
  \BibitemOpen
  \bibfield  {author} {\bibinfo {author} {\bibfnamefont {A.~G.}\ \bibnamefont
  {Volosniev}}, \bibinfo {author} {\bibfnamefont {N.~T.}\ \bibnamefont
  {Zinner}}, \bibinfo {author} {\bibfnamefont {D.~V.}\ \bibnamefont {Fedorov}},
  \bibinfo {author} {\bibfnamefont {A.~S.}\ \bibnamefont {Jensen}}, \ and\
  \bibinfo {author} {\bibfnamefont {B.}~\bibnamefont {Wunsch}},\ }\href
  {\doibase 10.1088/0953-4075/44/12/125301} {\bibfield  {journal} {\bibinfo
  {journal} {J. Phys. B}\ }\textbf {\bibinfo {volume} {44}},\ \bibinfo {pages}
  {125301} (\bibinfo {year} {2011}{\natexlab{b}})}\BibitemShut {NoStop}%
\bibitem [{\citenamefont {Boll{\' e}}\ and\ \citenamefont
  {Gesztesy}(1984)}]{BolGes84}%
  \BibitemOpen
  \bibfield  {author} {\bibinfo {author} {\bibfnamefont {D.}~\bibnamefont
  {Boll{\' e}}}\ and\ \bibinfo {author} {\bibfnamefont {F.}~\bibnamefont
  {Gesztesy}},\ }\href {\doibase 10.1103/PhysRevLett.52.1469} {\bibfield
  {journal} {\bibinfo  {journal} {Phys. Rev. Lett.}\ }\textbf {\bibinfo
  {volume} {52}},\ \bibinfo {pages} {1469} (\bibinfo {year}
  {1984})}\BibitemShut {NoStop}%
\bibitem [{\citenamefont {Simon}(1976)}]{Sim76}%
  \BibitemOpen
  \bibfield  {author} {\bibinfo {author} {\bibfnamefont {B.}~\bibnamefont
  {Simon}},\ }\href {\doibase 10.1016/0003-4916(76)90038-5} {\bibfield
  {journal} {\bibinfo  {journal} {Ann. Phys.}\ }\textbf {\bibinfo {volume}
  {97}},\ \bibinfo {pages} {279} (\bibinfo {year} {1976})}\BibitemShut
  {NoStop}%
\bibitem [{\citenamefont {Newton}(1982)}]{New82}%
  \BibitemOpen
  \bibfield  {author} {\bibinfo {author} {\bibfnamefont {R.~G.}\ \bibnamefont
  {Newton}},\ }\href@noop {} {\emph {\bibinfo {title} {Scattering Theory of
  Waves and Particles}}},\ \bibinfo {edition} {2nd}\ ed.\ (\bibinfo
  {publisher} {Springer},\ \bibinfo {address} {New York},\ \bibinfo {year}
  {1982})\BibitemShut {NoStop}%
\bibitem [{\citenamefont {Rakityansky}\ and\ \citenamefont
  {Sofianos}(1998)}]{RakSof98}%
  \BibitemOpen
  \bibfield  {author} {\bibinfo {author} {\bibfnamefont {S.~A.}\ \bibnamefont
  {Rakityansky}}\ and\ \bibinfo {author} {\bibfnamefont {S.~A.}\ \bibnamefont
  {Sofianos}},\ }\href {\doibase 10.1088/0305-4470/31/22/015} {\bibfield
  {journal} {\bibinfo  {journal} {J. Phys. A}\ }\textbf {\bibinfo {volume}
  {31}},\ \bibinfo {pages} {5149} (\bibinfo {year} {1998})}\BibitemShut
  {NoStop}%
\bibitem [{\citenamefont {Newton}(1986)}]{New86}%
  \BibitemOpen
  \bibfield  {author} {\bibinfo {author} {\bibfnamefont {R.~G.}\ \bibnamefont
  {Newton}},\ }\href {\doibase 10.1063/1.527294} {\bibfield  {journal}
  {\bibinfo  {journal} {J. Math. Phys.}\ }\textbf {\bibinfo {volume} {27}},\
  \bibinfo {pages} {2720} (\bibinfo {year} {1986})}\BibitemShut {NoStop}%
\bibitem [{\citenamefont {Pikovski}\ \emph {et~al.}(2010)\citenamefont
  {Pikovski}, \citenamefont {Klawunn}, \citenamefont {Shlyapnikov},\ and\
  \citenamefont {Santos}}]{PikKlaShl10}%
  \BibitemOpen
  \bibfield  {author} {\bibinfo {author} {\bibfnamefont {A.}~\bibnamefont
  {Pikovski}}, \bibinfo {author} {\bibfnamefont {M.}~\bibnamefont {Klawunn}},
  \bibinfo {author} {\bibfnamefont {G.~V.}\ \bibnamefont {Shlyapnikov}}, \ and\
  \bibinfo {author} {\bibfnamefont {L.}~\bibnamefont {Santos}},\ }\href
  {\doibase doi:10.1103/PhysRevLett.105.215302} {\bibfield  {journal} {\bibinfo
   {journal} {Phys. Rev. Lett.}\ }\textbf {\bibinfo {volume} {105}},\ \bibinfo
  {pages} {215302} (\bibinfo {year} {2010})}\BibitemShut {NoStop}%
\bibitem [{\citenamefont {Zinner}\ \emph {et~al.}(2010)\citenamefont {Zinner},
  \citenamefont {Wunsch}, \citenamefont {Pekker},\ and\ \citenamefont
  {Wang}}]{ZinWunPek10}%
  \BibitemOpen
  \bibfield  {author} {\bibinfo {author} {\bibfnamefont {N.~T.}\ \bibnamefont
  {Zinner}}, \bibinfo {author} {\bibfnamefont {B.}~\bibnamefont {Wunsch}},
  \bibinfo {author} {\bibfnamefont {D.}~\bibnamefont {Pekker}}, \ and\ \bibinfo
  {author} {\bibfnamefont {D.-W.}\ \bibnamefont {Wang}},\ }\href@noop {} {\
  (\bibinfo {year} {2010})},\ \Eprint {http://arxiv.org/abs/1009.2030}
  {arXiv:1009.2030} \BibitemShut {NoStop}%
\end{thebibliography}%
\end{document}